\documentclass[aps,prl,reprint,superscriptaddress]{revtex4-2}
\usepackage[utf8]{inputenc}
\usepackage[english]{babel}
\usepackage{amsmath}
\usepackage{amsfonts}
\usepackage{amssymb}
\usepackage{graphicx}
\usepackage{xcolor}
\usepackage[left=2cm,right=2cm,top=2cm,bottom=2cm]{geometry}
\usepackage{soul}
\usepackage{hyperref}

\newcommand{\expval}[1]{\langle #1 \rangle}

\newcommand{\pdagger}{\phantom{\dagger}}

\newcommand{\nsigma}{\overline{\sigma}}

\newcommand{\bonnpi}{Physikalisches Institut, University of Bonn, Nussallee 12, 53115 Bonn, Germany}
\newcommand{\salerno}{Physics Department ``E. R. Caianiello" and CNR-SPIN, Universitá degli Studi di Salerno, INFN, Gruppo Collegato di Salerno, 84084-Fisciano (Sa), Italy}
\newcommand{\lyon}{ENSL, CNRS, Laboratoire de Physique, F-69342 Lyon, France}

\begin{document}

\title{Superconductivity in the repulsive Hubbard model on different geometries induced by density-assisted hopping}

\author{Franco T. Lisandrini}
\affiliation{\bonnpi}
\author{Edmond Orignac}
\affiliation{\lyon}
\author{Roberta Citro}
\affiliation{\salerno}
\author{Ameneh Sheikhan}
\affiliation{\bonnpi}
\author{Corinna Kollath}
\affiliation{\bonnpi}

\begin{abstract}
We study the effect of density-assisted hopping  on different
dimerized lattice geometries, such as bilayers and ladder structures. We show analytically that the density-assisted
hopping induces an attractive interaction in the lower
(bonding) band of the dimer structure and a repulsion in the upper (anti-bonding) band. Overcoming the onsite repulsion, this can lead to the appearance of superconductivity with a pairing structure more complex than s-wave pairing.
Combining numerical and analytical
methods such as the matrix product states ansatz, bosonization and perturbative calculations we map out the phase diagram of the two-leg ladder system and identify its superconducting phase. We characterize the transition from the non-density-assisted repulsive regime to the spin-gapped superconducting regime as a Berezinskii-Kosterlitz-Thouless transition.
\end{abstract}

\maketitle

%%%%%%%%%%%%%%%%%%%%%%%%%%%%%%%%%%%%
% \section*{Introduction}
%%%%%%%%%%%%%%%%%%%%%%%%%%%%%%%%%%%%

\begin{figure}
\centering
\includegraphics[width=\columnwidth]{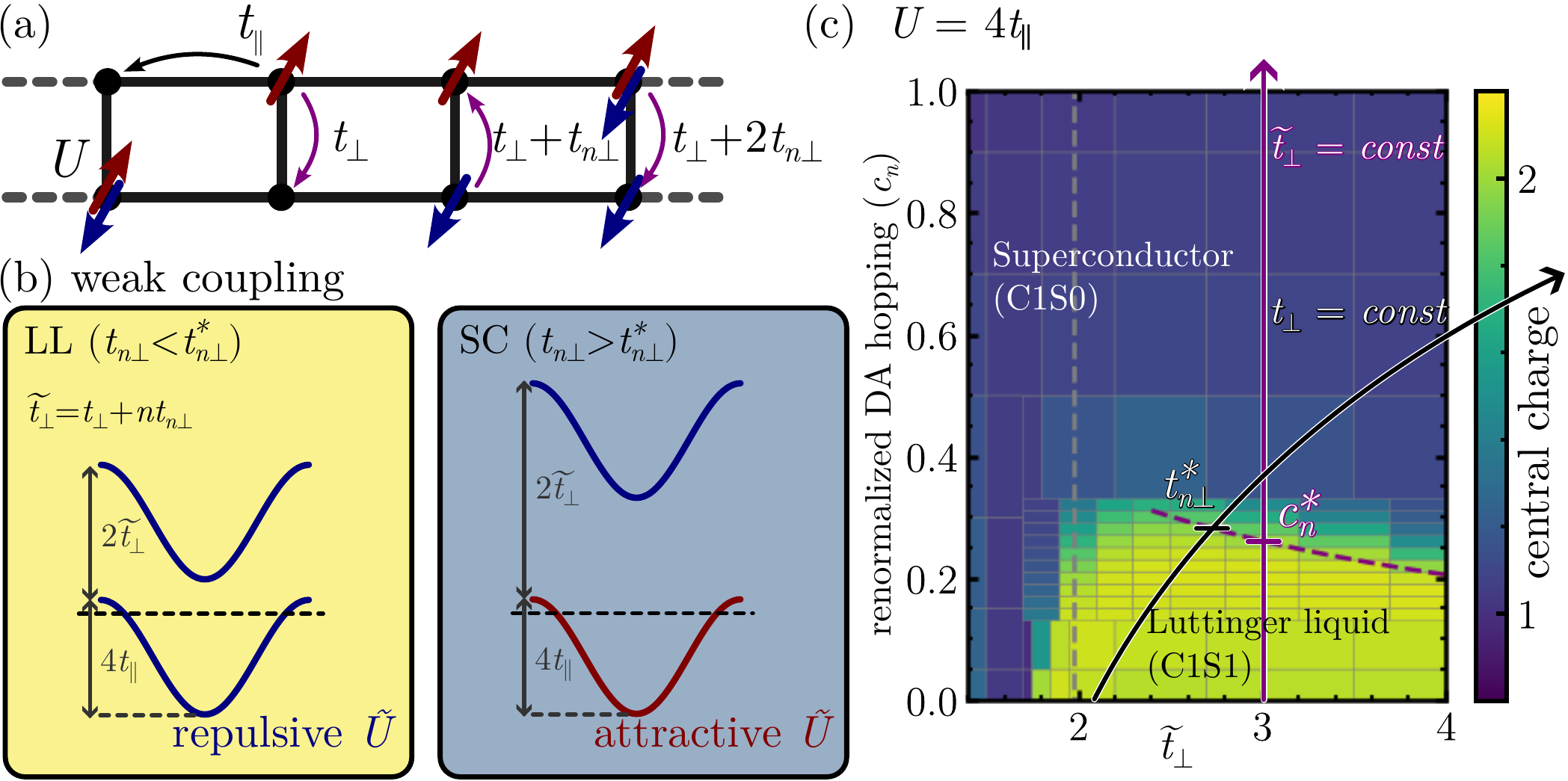} 
\caption{ (a) Sketch of the Hubbard ladder with $L$ rungs. The value of the on-site interaction is $U$, hopping amplitudes on different directions are $t_\parallel$ and $t_\perp$, the density-assisted hopping contribution is $t_{n\perp}$. (b) In the weak coupling limit, the density assisted hopping term makes the bonding band attractive;  (c) the value of the central charge extracted from MPS simulations in the $U=4t_\parallel$ case. The green (purple) line shows an example of a constant (effective) perpendicular hopping $t_{\perp}$ ($\tilde{t}_{\perp}$) level line. The effective attraction opens a spin gap and induces a superconducting phase. The corresponding simulation parameters are $n=0.9375$, $U=4t_{\parallel}$, and $L=80$ rungs. 
\label{FIG_system}}
\end{figure}

The theoretical description of high critical temperature superconductivity is a central subject in condensed matter physics. 
To tackle this problem, a commonly used model to try to describe the physics of cuprates is the two-dimensional Hubbard model \cite{LeeWen2006, Scalapino2012, ArovasRaghu2022, LiuChan2025, XiaoZhang2023, QinGull2022}. 
This is a very simple model with just two competing terms, yet it hosts a very rich phase diagram. 
Despite this very rich and complex phase diagram, it is still under investigation whether superconductivity occurs in the plain repulsive Hubbard model \cite{QinZhang2020, QinGull2022}. 
However, the addition of diagonal hopping $t'$, helps tip the balance in favor of superconductivity \cite{ZhengChan2017, QinGull2022, XuZhang2024, ZhangvonDelft2025}. 
Many extensions of the Hubbard model have been studied in the search for more realistic models of high-temperature superconductors \cite{JiangDevereaux2019, PonsioenCorboz2019, JiangJiang2020, JiangKivelson2021}.
These include density-assisted hopping terms \cite{Hirsch1989, HirschMarsiglio1989, SimonAligia1993,JaparidzeMuellerHartmann1994, JaparidzeKampf1999, ArracheaAligia1999, ArracheaAligia2000, AligiaTorio2007, DobryAligia2011, AligiaArrachea1999, JiangWhite2023, Controzzi1999}, which arise naturally from the down-folding of the three-band Hubbard model describing the Cu-O planes in cuprates\cite{VarmaAbrahams1987, Emery1987} to a single-band Hubbard model. 
They were studied in the context of two-dimensional hole superconductivity to explain the asymmetry between particle and hole doping \cite{Hirsch1989, HirschMarsiglio1989}. Recently it has been shown numerically \cite{JiangWhite2023} that the amplitude of the density-assisted hopping term in cuprates is 60\% of that of the normal hopping and they can increase the mobility of pairs of holes beyond mean-field effects. 
The inclusion of density-assisted hopping terms in the Hubbard model is motivated not only by solid-state systems, but also by possible direct experimental realizations in ultracold-atom experiments \cite{JuergensenLuehmann2014, DuttaZakrzewski2015, MeinertNaegerl2016, KlemmerBergschneider2024, Eckardt2017, GoergEsslinger2019, ChandaZakrzewski2025}. 

We consider dimers formed by two sites with Hubbard repulsion. The two sites within the dimers are coupled by normal and density-assisted hopping terms. The inter-dimer coupling is a normal hopping. 
Examples of dimer lattices in one and two dimensions are the Hubbard ladder \cite{NoackScalapino1994, NoackScalapino1995, NoackScalapino1996, NoackZacher1997, DolfiTroyer2015, ShenQin2023, SheikhanKollath2020, BalentsFisher1996, Giamarchibook, GannotKivelson2020} and the bilayer Hubbard model \cite{LanataFabrizio2009, MaierScalapino2011, GallKoehl2021}. 
In the ladder geometry [see Fig.~\ref{FIG_system}(a)], the two chains are connected by normal and density-assisted hopping along the rungs, whereas in the bilayer, the two layers are connected this way. 

We show that density-assisted hopping can induce a superconducting phase (blue region in Fig.~\ref{FIG_system}c) over an originally normal phase (yellow region in Fig.~\ref{FIG_system}c). The origin of the superconductivity lies in an effective attractive interaction which is induced for a range of fillings by the density-assisted hopping (see Fig.~\ref{FIG_system}b). 
This  mapping to an effectively attractive model is independent of the underlying lattice geometry and in particular also holds in the bilayer system. 
The amplitude of density-assisted hopping needed to induce superconductivity in our system is similar to the amplitude found for these terms in cuprates \cite{JiangWhite2023}. 
We confirm our analytical results for the geometry of a ladder by the results from the numerically exact Matrix Product States (MPS) method (i.e., the density matrix renormalization group algorithm). 
We identify the effectively attractive interaction as the new pairing mechanism which is the origin of the induced superconducting correlations and thus the extended superconducting phase space. 

%%%%%%%%%%%%%%%%%%%%%%%%%%%%%%%%%%%%
% \section*{Model}
%%%%%%%%%%%%%%%%%%%%%%%%%%%%%%%%%%%%

\textit{Extended Hubbard model.--}
The Fermi-Hubbard model on a dimerized lattice is given by $H_\text{FH} = H_\parallel + H_\perp + H_U $, where the hopping between the lattice sites is
$
H_\parallel =- t_\parallel 
\sum_{\langle i,j \rangle,\alpha,\sigma} \left( \hat{c}_{\alpha,\sigma,i}^{\dagger} \hat{c}_{\alpha,\sigma,j}^{\pdagger} + \text{H.c.} \right) \nonumber $, the hopping inside the dimers is 
$
 H_\perp=- t_\perp \sum_{j,\sigma} \left( \hat{c}_{1,\sigma,j}^{\dagger} \hat{c}_{2,\sigma,j}^{\pdagger} + \text{H.c.} \right)$, and the onsite interaction strength is $
 H_U=\, U \sum_{j,\alpha} \hat{n}_{\alpha,\uparrow,j} \hat{n}_{\alpha, \downarrow, j}$.
Here, $\hat{c}_{\alpha,\sigma,j}$ annihilates a fermion of spin $\sigma=\uparrow, \downarrow$ at dimer $j$ and $\alpha=1,2$ labeling the internal degree of freedom of the dimer. The summation index $\langle i,j \rangle$ indicates the nearest neighbors between dimers (in-layer hopping for bilayers, chain hopping for the ladders). 
The number operator is denoted by $\hat{n}_{\alpha,\sigma,j}$ = $\hat{c}_{\alpha,\sigma,j}^{\dagger} \hat{c}_{\alpha,\sigma,j}^{\pdagger}$, and the density is $ n = N/2L$ where $N$ is the number of fermions and $L$ is the number of dimers.

To obtain insights for large coupling of the dimers, it is useful to perform the transformation $\hat{c}_{ k_\perp, \sigma, j} = \left( \hat{c}_{1, \sigma, j} \pm \hat{c}_{2, \sigma, j} \right) /\sqrt{2}, $
where the index $k_\perp=0,\pi$ corresponds to bands hereafter referred to as the bonding and antibonding bands, respectively. 
This transformation makes the perpendicular hopping term diagonal, 
 $H_{\perp} = 
 - t_{\perp} \sum_{\sigma,j}
 \left( \hat{n}_{  0, \sigma, j} - \hat{n}_{\pi, \sigma, j} \right).$
For the non-interacting case ($U=0$), we find two independent bands shifted by an energy $2t_\perp$ where the bonding (antibonding) band is lower (higher) in energy (see sketch in Fig.~\ref{FIG_system}b). 
For the remainder of this work, we will focus mainly on the regime where the two bands are well separated. 
For example, in the non-interacting ($U=0$) one-dimensional system below half filling ($n<1$), both bands are partially occupied for $t_\perp <  2t_\parallel | \cos (\pi n)| $, while for $t_\perp > 2t_\parallel | \cos (\pi n)| $, only the bonding band remains partially filled due to increased band separation. 

We introduce a density-assisted tunneling term inside the dimers, where the amplitude of hopping becomes dependent on the local densities of the opposite spin on the dimer, i.e., 
\begin{equation*}
 H_\text{DA} = - t_{n\perp} \sum_{\sigma,j} 
 \left( \hat{n}_{1,\nsigma,j} +  \hat{n}_{2,\nsigma,j} \right)
 \left( \hat{c}_{1,\sigma,j}^{\dagger} \hat{c}_{2,\sigma,j}^{\pdagger} + \text{H.c.} \right)
\end{equation*}
The extended Hamiltonian includes the density-assisted hopping $H=H_\text{FH}+H_\text{DA}$. 
At the mean-field level, where the local densities are approximated as $ \hat{n}_{\alpha,\sigma,j} \simeq n/2 $, the density-assisted term modifies the rung hopping amplitude \cite{Controzzi1999}. As a result, the effective hopping inside a dimer becomes 
\begin{equation*}
\tilde{t}_\perp = t_\perp + n t_{n\perp}
\end{equation*}
and thus the energy difference between the bands is $2\tilde{t}_\perp$. 
We call 
\begin{equation*}
c_{n} = nt_{n\perp} / \tilde{t}_\perp
\end{equation*}
the density-assisted hopping ratio, since it continuously interpolates between the non-density-assisted case ($t_{\perp}=\tilde{t}_\perp$ and $t_{n\perp}=0$ for $c_{n}=0$) and the only-density-assisted case ($t_{\perp}=0$ and $t_{n\perp}=\tilde{t}_\perp /n $ for $c_{n}=1$). 

\begin{figure*}
\centering
\includegraphics[width=0.80\columnwidth]{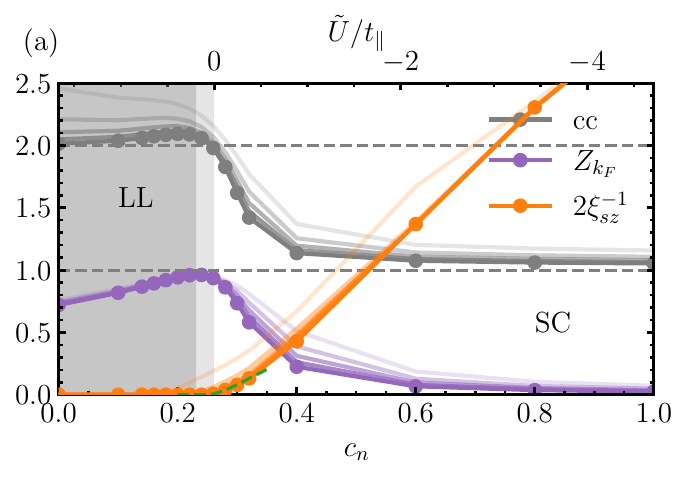} 
\includegraphics[width=0.45\columnwidth]{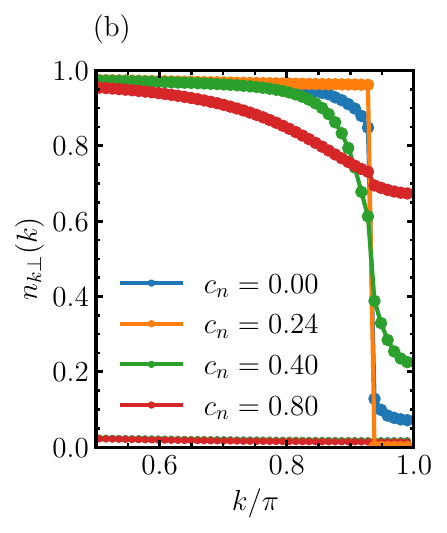} 
\includegraphics[width=0.80\columnwidth]{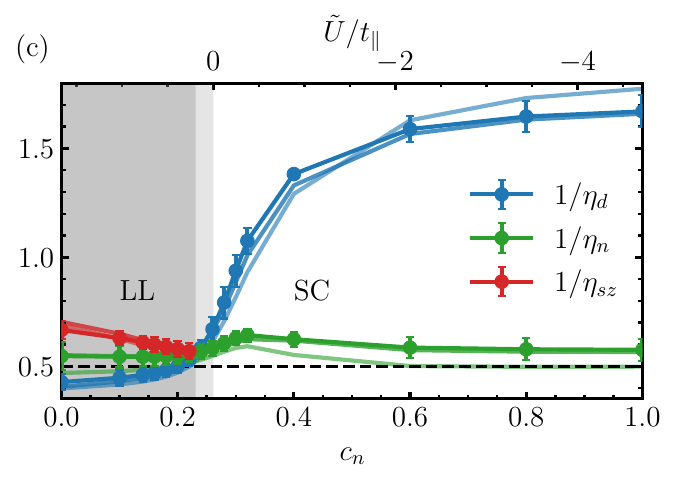} 
\caption{ Transition at $U=4t_\parallel$, $n=0.9375$ and $\tilde{t}_\perp=3t_\parallel$ for a $96$-rung ladder. (a) The gray and purple plots correspond to the central charge and the discontinuity of the momentum distribution $Z_{k_F}$, respectively. The orange plot corresponds to the inverse correlation length. 
(b) Momentum distribution for the up-spin in the bonding and antibonding band, $n(k)$, for different values of the density-assisted hopping ratio ($c_n$). The occupation of the antibonding band stays below 3\%. 
(c) Power-law exponents for the pair ($\eta_d$), density ($\eta_n$) and spin ($\eta_{sz}$) correlation functions. The plots are $1/\eta$, hence a larger value means than the correlations decay more slowly. The error bars were estimated by changing the fitting range and obtaining the maximum deviation. With translucid lines we plot the results for $L=32, 48, 64, 80$ in panel (a), and $L=64, 80$ in (c) (more opaque lines correspond to larger systems). 
\label{FIG_cc_gap_Zk}}
\end{figure*}

To understand how the interaction and the density-assisted process couple the bonding and antibonding bands, we represent the interaction Hamiltonian $H_U$ and the density-assisted Hamiltonian $H_\text{DA}$ in the bonding/antibonding basis. 
The density-assisted hopping term reads: 
\begin{equation*}
H_\text{DA} = - 2t_{n\perp} \sum_{j} 
 \left( n_{  0, \uparrow, j} n_{  0, \downarrow, j} - n_{\pi, \uparrow, j} n_{\pi, \downarrow, j} \right).
\end{equation*}
Remarkably, these terms act exclusively as interactions within each band. Moreover, when $t_{n\perp} > 0$, the interaction is attractive in the bonding band ($k_\perp=0$) while it is repulsive in the upper band ($k_\perp=\pi$). 

In the new basis, the interaction Hamiltonian takes the more complicated form,
\begin{align}
H_U = \frac{U}{2} \sum_{j} 
&\left\{\left( 
  \hat{n}_{  0,\uparrow, j} \hat{n}_{  0, \downarrow, j}
+ \hat{n}_{\pi, \uparrow, j} \hat{n}_{\pi, \downarrow, j}
\right)\right. + \nonumber \\ 
  &\left( 
\frac{1}{2} \hat{n}_{0,j} \hat{n}_{\pi, j}
- 2\hat{\bf S}_{0, j} \hat{\bf S}_{\pi, j}
\right) + \nonumber \\ 
  &\left.\left(      
  \hat{c}_{  0, \uparrow, j}^{\dagger} \hat{c}_{  0, \downarrow, j}^{\dagger} 
  \hat{c}_{\pi, \downarrow, j}^{\pdagger} \hat{c}_{\pi, \uparrow, j}^{\pdagger} 
 + \text{H.c.} \right)\right\}, 
\label{EQ_U_AB}
\end{align}
where the spin operators are defined as $ \hat{\bf S}_{k_\perp, j} = \frac{1}{2} \sum_{\sigma \sigma'} \hat{c}_{k_\perp, \sigma, j}^{\dagger} {\boldsymbol \sigma}_{\sigma \sigma'} \hat{c}_{k_\perp, \sigma', j}^{\pdagger} $, and ${\boldsymbol \sigma}$ being the vector of Pauli matrices. 
The local density operators in each band are summed over both spins as $\hat{n}_{k_\perp j} =  \hat{n}_{ k_\perp,\uparrow, j}+\hat{n}_{ k_\perp, \downarrow, j}$. 
Repulsion between particles in the original basis transforms into intraband repulsion, with an interaction strength $U/2$, reduced by half of its original value. In addition, the transformation generates interband terms. 
The inter-band density-density interaction penalizes the simultaneous occupation of both sites in the dimer, effectively penalizing the occupation of the upper band. Naively, one could expect the inter-band ferromagnetic spin-spin interaction term to favor the occupation of the upper band, but even for the rung triplet state, the energy gain by the spin interaction does not overcome the energy loss from the density interaction. 

%%%%%%%%%%%%%%%%%%%%%%%%%%%%%%%%%%%%
% \section*{DAH as an attraction}
%%%%%%%%%%%%%%%%%%%%%%%%%%%%%%%%%%%%

Consequently, we study the system in the regime where the upper band has low occupation $\expval{\hat{n}_{\pi j}} \approx 0$ and $\expval{ \hat{\bf S}_{\pi j} } \approx {\bf 0}$, therefore, the spin and density interactions between the bands can be safely neglected (second line of Eq.~\ref{EQ_U_AB}). Pair hopping (third line of Eq.~\ref{EQ_U_AB}) is the only remaining term that couples the two bands. 
Considering this process as a perturbation up to the second order, we are able to effectively decouple the two bands. 
The resulting effective Hamiltonian describing the bonding band is 
\begin{equation}
\tilde{H} = \tilde{U} \sum_{j} \hat{n}_{0,\uparrow,j} \hat{n}_{0, \downarrow, j} 
 - t_\parallel \sum_{\sigma,\langle i,j \rangle} \left( \hat{c}_{0,\sigma,j}^{\dagger} \hat{c}_{0,\sigma,j+1}^{\pdagger} + \text{H.c.} \right),
 \label{EQ_Heff}
\end{equation}
%where $\tilde{U} = U/2 - 2 t_{n\perp}  - (U/2)^2 / \Delta E$, and $\Delta E = 4t_\perp + 4t_{n\perp}$. 
where $\tilde{U} = U/2 - 2c_{n} \tilde{t}_\perp / n  - (U/2)^2 / \Delta E$, and $\Delta E = 4t_\perp + 4t_{n\perp}$. 
Since this expression is perturbative in $U/\Delta E$, Eq.~\ref{EQ_Heff} reduces to the repulsive single-band Fermi–Hubbard model on the underlying lattice geometry when there is no density-assisted hopping ($c_n=0$). 
However, if $c_n$ is large enough, the repulsive Hubbard model becomes an attractive Hubbard model in the bonding band. This effectively attractive Hubbard model can then host superconducting phases. The effective attraction in the bonding band is independent of the underlying lattice geometry and applies in particular to the ladder and the bilayer geometry. The formed pairs correspond to s-wave pairs within the bonding basis and thus have a richer structure in the original basis.
Annihilating an $s$-wave pair in the bonding band ($\hat{c}_{0, \uparrow, j} \hat{c}_{0, \downarrow, j}$) annihilates a superposition of on-site pairs ($s$-wave, i.e., $\Delta_{s, j} = \hat{c}_{1, \uparrow, j} \hat{c}_{1, \downarrow, j} + \hat{c}_{2, \uparrow, j} \hat{c}_{2, \downarrow, j}$) and dimer singlets (sometimes called $d$- or $p$-wave pairs in the ladder geometry, i.e., $\hat{\Delta}_{d, j} = \hat{c}_{1, \uparrow, j} \hat{c}_{2, \downarrow, j} - \hat{c}_{1, \downarrow, j} \hat{c}_{2, \uparrow, j}$). 
Therefore, the effective attraction introduced by density-assisted hopping not only promotes pair formation but also favors the emergence of non-local pairing components.

In the following, we focus on the extended Hubbard model on the ladder geometry.
The Fermi-Hubbard ladder provides a powerful platform to explore strongly correlated fermions in quasi-one-dimensional systems. 
In particular, close to half-filling, the Fermi-Hubbard ladder was the center of attention for the community for many years, both numerically \cite{NoackScalapino1994, NoackScalapino1995, NoackScalapino1996, NoackZacher1997, DolfiTroyer2015, ShenQin2023, SheikhanKollath2020} and analytically \cite{BalentsFisher1996, Giamarchibook, GannotKivelson2020}. 
It is a prototypical case of a system that presents superconducting correlations even in the presence of strong repulsive interactions. 

We start analyzing the effective Hamiltonian $\tilde{H}$, which in the ladder geometry represents a Hubbard chain. 
In the repulsive regime, bosonization predicts a Luttinger liquid phase with two gapless modes and a central charge from a conformal field theory of $c = 2$. This phase is normally dubbed C1S1, where the notation CXSY implies that the system hosts X gapless charge modes and Y gapless spin modes. 
Increasing the density-assisted hopping ratio $c_n$ transforms the initially repulsive interaction in the bonding band into an attractive one (see Fig.~\ref{FIG_system}b). 
For attractive interactions, bosonization calculations in the weak-coupling regime predict the opening of a spin gap \cite{Giamarchibook}. As a result, the system undergoes a transition from the gapless C1S1 phase to a spin-gapped C1S0 phase normally known as Luther-Emery phase \cite{LutherEmery1974, Giamarchibook}. 
This transition takes place at a critical hopping ratio which we will call $c_n^{\star}$. 
The spin gap causes the spin correlations to decay exponentially to zero with a finite correlation length whose magnitude is inversely proportional to the size of the gap ($\Delta_{sz} \propto \xi_{sz}^{-1}$) \cite{Giamarchibook}. 

The analytical results obtained from bosonization provide valuable insights into the low-energy physics of the system in the weakly interacting regime. However, to explore whether these predictions also hold in the strongly interacting limit, we complement our analysis with numerical simulations using MPS methods for the full Hamiltonian $H$ in the ladder geometry. 
The simulations in this work were made using the Itensor libraries \cite{FishmanStoudenmire2022, FishmanStoudenmire2022_code}. 
We calculate three observables to probe the quantum phase transition: the central charge, the spin correlation length, and the discontinuity of the momentum distribution at the Fermi momentum [Fig.~\ref{FIG_cc_gap_Zk} (a)], complemented by the analysis of the density, pair, and spin correlation functions [Fig.~\ref{FIG_cc_gap_Zk} (c)]. In our MPS calculations, we consider a relatively strong interaction $U=4t_\parallel$, a filling $n=0.9375$ corresponding to one hole per sixteen rungs, and unless otherwise stated, the effective rung-hopping amplitude is $\tilde{t}_\perp = 3t_\parallel$. The critical value for the density-assisted hopping ratio predicted by Eq.~\ref{EQ_Heff}, at which the effective interaction in the bonding band changes sign, is $c_n^{\star} \simeq 0.26$. 
We have kept a maximum of $\chi=5000$ states during the MPS truncation. 
The resulting truncation error strongly depends on the effective rung hopping $\tilde{t}_\perp$. For $\tilde{t}_\perp = 3t_\parallel$, we have obtained a truncation error below $10^{-10}$ and well-converged results. 
With decreasing rung hopping the calculations become more demanding, and therefore we see an increase on the truncation error, e.g.~$10^{-8}$ for $t_\perp=1.8t_{\parallel}$. 

%%%%%%%%%%%%%%%%%%%%%%%%%%%%%%%%%%%%
% Central charge
We extract the central charge [see Fig.~\ref{FIG_cc_gap_Zk}(a) (gray dots)] from the scaling of the entanglement entropy with subsystem size, i.e., $S_\text{vN}(l) = A + B t(l) + c \ln d(l|L) /6 $, where $d(l|L) = 2L | \sin(\pi l /L) |/\pi$ is the conformal distance and $t(l)= \sum_{\alpha,\sigma} \mathrm{Re} \expval{\hat{c}_{\alpha,\sigma,l}^{\dagger} \hat{c}_{\alpha,\sigma,l+1}^{\pdagger}}$ is the local kinetic energy (an example is given in Fig.~\ref{FIG_ccfit} of the supplemental material \cite{LisandriniKollath2026_SM}). 
As expected, the central charge decreases from $c \simeq 2$ in the C1S1 phase to $c \simeq 1$ in the C1S0 phase [Fig.~\ref{FIG_cc_gap_Zk}(a)]. 
We attribute the deviations from the exact value to finite-size effects [different curve transparencies indicating the system size in Fig.~\ref{FIG_cc_gap_Zk}(a)]. With increasing system size, one should recover a sharp step \cite{LaeuchliKollath2008}. 
 
The strong finite-size effect arises from a contribution to the central charge from the gapped mode, which decreases as the system size increases. (for more details on this, see the supplemental material \cite{LisandriniKollath2026_SM}). For this reason, we obtain a monotonic decreasing behavior of the extracted value of the central charge with the size of the system.  
Therefore, once the central charge is below $c=2$, we can say that we are above the transition. Following this argument, one can estimate from the behaviour of the central charge that the transition lies below $c_n < 0.25$, which is close but below the analytically predicted value ($c_n^{\star} \simeq 0.26$). 

%%%%%%%%%%%%%%%%%%%%%%%%%%%%%%%%%%%%
% Gap
To further characterize the C1S1–C1S0 transition, in Fig.~\ref{FIG_cc_gap_Zk}(a), we study the behavior of the inverse of the correlation length $\xi_{sz}^{-1}$. 
The spin correlation length was extracted from the spin correlation function by fitting $A e^{-l/\xi_{sz}} l^{-\eta_{sz}}$ (with $A$, $\xi_{sz}$ and $\eta_{sz}$ as fitting parameters, for more details see the supplemental material \cite{LisandriniKollath2026_SM}). 
As mentioned above, the correlation length is inversely proportional to the gap. 
We see that, at low values of the density-assisted hopping ratio $c_n$, the gap vanishes within numerical accuracy. This is in agreement with the C1S1 phase. When increasing $c_n$, a spin gap slowly opens, agreeing with a transition to the C1S0 phase. 
The green dashed line corresponds to a fitting of the function $\Delta_{sz}\propto \xi_{sz} ^{-1}= a \exp( b / \sqrt{c_n - c_n^{\ast}} )$, with $a$, $b$ and $c_n^{\ast}$ as fitting parameters. 
The performed fit is consistent with the Berezinskii–Kosterlitz–Thouless (BKT) transition\cite{Berezinskii1972, KosterlitzThouless1973, Kosterlitz1974, Giamarchibook, BenfattoGiamarchi2013, BouchouleWeber2025} from the gapless C1S1 phase to the spin-gapped C1S0 phase. 
This type of transition is characterized by an essential singularity, where the spin gap opens exponentially slowly near the critical point.  
The value of the transition extracted from the fitting is approximately $c_n^{\ast 1} = 0.24(1)$, which is in agreement with the value $c_n < 0.25$ extracted from the behaviour of the central charge.
The uncertainty was taken to be half the distance between data points. 

%%%%%%%%%%%%%%%%%%%%%%%%%%%%%%%%%%%%
% Single-particle momentum distribution
In order to test the sign change of the effective interaction in the bonding band obtained analytically, we numerically calculate the momentum distribution. It is obtained from the Fourier transform of the single-particle correlation function for the up-spin fermions, i.e.,  $n_{k\perp}(k) = \text{FT}[ \expval{ \hat{c}^{\dagger}_{k\perp, \uparrow, i} \hat{c}^{\pdagger}_{k\perp, \uparrow, j} }_c]$, where the subindex $c$ denotes the connected correlation function (see supplemental material for more details \cite{LisandriniKollath2026_SM}) and $k_\perp = 0, \pi$ refers to the bonding and antibonding band, respectively. We present the results for different density-assisted hopping ratio values in Fig.~\ref{FIG_cc_gap_Zk}(b). For our calculations at $\tilde{t}_\perp = 3t_\parallel$, the antibonding band (small circles) is nearly empty with less than 3\% of all fermions. However, the bonding band (big circles) represents approximately 97\% of the particles. This observation is consistent with the picture of two well-separated bands, where only the bonding band is partially filled. 
We observe a jump at the Fermi momentum $k_F$ in the momentum distribution for all curves in Fig.~\ref{FIG_cc_gap_Zk}(b) due to finite size effects. This allows us to define the momentum discontinuity as $Z_{k_F} = \max_{j}[n(k_j)-n(k_{j+1})].$ 
Let us note that the jump for most parameters is due to finite-size effects, since in the thermodynamic limit and in the presence of interactions, $Z_{k_F}$ vanishes in one-dimensional systems due to the absence of well-defined quasiparticles. 
However, at the non-interacting point ($\tilde{U}=0$), i.e., at the transition point $c_n^{\star}$, for the effective one-band model (Eq.~\ref{EQ_Heff}), the momentum distribution should be a step function and the value of $Z_{k_F}$ should reach unity. In Fig.~\ref{FIG_cc_gap_Zk}(b), we show with orange lines the momentum distribution close to the critical point at $c_n = 0.24$. It behaves as a sharp step function with $Z_{k_F} \simeq 0.97$. 
Furthermore, as shown in Fig.~\ref{FIG_cc_gap_Zk}(a), $Z_{k_F}$ is not affected by system size at this point. We attribute the deviation from a step of $1$ to the remaining interactions with the antibonding band which has a very small occupation. 
The analytical picture of a sign change of the effective interaction is confirmed by our MPS results for $Z_{k_F}$ [see Fig.~\ref{FIG_cc_gap_Zk}(a) (purple dots)]. The maximum of the momentum discontinuity $Z_{k_F}$ is estimated at $c_n^{*2} = 0.23(1)$, agreeing with the value extracted from the gap opening [$c_n^{*1} = 0.24(1)$]. 
The precise scaling behavior of the momentum discontinuity $Z_{k_F}$ away from the critical point agrees with bosonization predictions, and the asymmetry with respect to this point is a consequence of different scaling exponents on either side (see supplemental material for more details \cite{LisandriniKollath2026_SM}). 

%%%%%%%%%%%%%%%%%%%%%%%%%%%%%%%%%%%%
% Exponents
As previously outlined, the density, pair, and spin correlation functions are fundamental to characterizing the system's phases. 
The precise definitions of these correlation functions are 
$ C_{n}(i,j) = \expval{ \hat{n}_{0, j} \hat{n}_{0, i} }_c$, 
$ C_{sz}(i,j) = \expval{ \hat{S}^z_{0, j} \hat{S}^z_{0, i} }_c$, and
$ C_{d}(i,j) = \expval{ \hat{\Delta}^{\dagger}_{d, i} \hat{\Delta}^{}_{d, j} }_c$. 
The operator $\hat{S}^z_{k_\perp, j}$ is the $z$-component of the spin operator defined below Eq.~\ref{EQ_U_AB}. 
Note that these were chosen to be symmetric under leg inversion and therefore acting on the symmetry sector of the bonding band. 
In the C1S0 phase (Luther-Emery), bosonization predicts that the long-distance behavior of the density and pair correlation functions is governed by a power-law decay determined by the Luttinger liquid parameter $K_\rho$ which can be modulated by an oscillatory term (see supplemental material \cite{LisandriniKollath2026_SM}). In this regime, the spin correlation function $C_{sz}$ decays exponentially. 
In the C1S1 phase (Luttinger liquid), the spin correlation function should also decay as a power law.

The exponents $\eta$ of the power-law term for each correlation function were extracted by fitting the function 
$A l^{-\eta} + B l^{-\eta} \cos(2\pi n_0 l)$
for the long-range correlations with $A$, $B$, and $\eta$ as fitting parameters and $n_0$ the average density at the center of the chain obtained from the density mean values (see the supplemental material for more details on the fitting procedure \cite{LisandriniKollath2026_SM}). 

The extracted exponents for the spin, density, and pair correlations are denoted as $\eta_{sz}$, $\eta_n$ and $\eta_d$, respectively, and their inverse were plotted in Fig.~\ref{FIG_cc_gap_Zk}(c). 
The correlation function with the largest inverse exponent decays most slowly and thus dominates at long distances, providing a signature of the underlying phase. 
As expected, in the C1S1 phase, the density correlations dominate over the pair correlation, but above the transition to the C1S0 phase, the pair correlations take over and dominate over the density correlations. 
The spin exponent $\eta_{sz}$ is not plotted above the transition because the spin correlations decay exponentially in this phase. 

For low values of density-assisted hopping ratio $c_n$ [spin-gapless (C1S1) phase], we observe that the three exponents for spin, density, and pairs remain close to $\eta^{-1} \sim 0.5$. 
Among them, the pair correlations is the least dominant while the spin correlations appear to be slightly more dominant than the density correlations. 
These differences lie within our numerical resolution. 
Analytically, the long-distance decay of spin and density correlations is expected to be governed by the same Luttinger liquid exponent. Additionally, the bosonization predicts the relation $\eta_n \eta_d=1$ for long distances.
We have studied the deviations from the expected behavior in the supplemental material and show that we can attribute them to finite-size effects \cite{LisandriniKollath2026_SM}. 

For intermediate-to-large values of the density-assisted hopping ratio, the value of the inverse exponent associated with $d$-wave pair correlations increases rapidly, while the exponent corresponding to density correlations does not change significantly. 
The difference between the exponents increases well-beyond finite-size effects or the uncertainty coming from our fitting procedure. 
This shows the emergence of a very robust unconventional superconducting phase, where the pair correlations are considerably larger and decay more slowly than the other correlations in the system. 

At the transition point, i.e., the non-interacting point, all three exponents are expected to tend to the same value $\eta^{-1} = 0.5$. 
We see this crossing in our data, up to small deviations from this predicted value of the order of the error bars. These are caused by finite system size and the limitations this imposes on our fitting procedure. 
We have performed runs on larger systems up to $L_x=112$ for selected points in the parameter space, and showed that the difference between the exponent extracted for $L_x=96$ and $L_x=112$ is below 1\% for $c_n=0.6$ and 0.1\% for $c_n=0.1$, well below the error bars indicated in Fig.~\ref{FIG_cc_gap_Zk}(c). 
The value of the density-assisted hopping ratio $c_n$ where the crossing takes place is $c_n^{\ast 3} = 0.23(1)$. 
Again, this agrees well with the values obtained from the gap opening ($c_n^{\ast 1}$) and the maximum discontinuity $Z_{k_F}$ ($c_n^{\ast 2}$) up to the uncertainty arising from the discretization of our data points. 
It is also very remarkable that the numerically extracted value for the $U=4t_\parallel$ case is very close to the critical value predicted from Eq.~\ref{EQ_Heff} for weak interactions ($c_n^{\star} \simeq 0.26$). 

The same transition we discussed here for a specific value of the effective hopping ($\tilde{t}_\perp = 3t_\parallel$) can be observed for a broader range of parameters. This is shown in Fig.~\ref{FIG_system}(c) where we use the central charge to map out the phase diagram. The transition from the C1S1 Luttinger liquid phase to the C1S0 unconventional superconducting phase induced by the density-assisted hopping occurs over a wide parameter regime. The critical coupling predicted from Eq.~\ref{EQ_Heff} gives a good approximation for the observed transition (see purple dashed line in Fig.~\ref{FIG_system}c). 

Thus, we have seen that in the dimerized ladder system, the density-assisted hopping induces superconductivity. Remarkably, the values at which the system becomes superconducting are similar to the values of density-assisted hopping estimated for cuprates \cite{JiangWhite2023}. 
This transition is very robust, and we find its signatures in a much broader parameter regime than the stringent validity regime of the effective model (see also the supplemental material for higher interaction strength and other fillings \cite{LisandriniKollath2026_SM}).

The results we obtained are not restricted to the case of the ladder geometry. 
In the cases of dimerized structures in two and three dimensions, the corresponding effective models are the two- and three-dimensional Hubbard models, respectively. 
These models are known to host a superconducting phase for attractive interactions \cite{QinGull2022, ScalettarDagotto1989, GunnarssonToschi2015, KellerSchollwoeck2001, ToschiCastellani2005, BauerDupuis2009}. Thus, also in higher dimensions, the density-assisted hopping term induces a superconducting 
phase at weak coupling. As in the one-dimensional setup, we expect that the superconducting phase exists to larger interaction strength much beyond the assumed validity regime of the effective model. 

%%%%%%%%%%%%%%%%%%%%%%%%%%%%%%%%%%%%

\textit{Summary and outlook.--}
We have studied the repulsive Hubbard model with density-assisted hopping on general geometries of dimerized lattices. 
We show that density-assisted hopping can induce interactions with a band-selective sign change. 
In the lower band, it generates an effective attraction between holes that favors the formation of dimer-singlet pairs with a structure more complex than conventional $s$-wave pairing. 
The resulting effective attraction therefore arises from a mechanism distinct from previously identified pairing mechanisms in Hubbard systems, and provides a route to emergent multiband superconductivity driven by correlated hopping.
We have presented strong numerical evidence that this picture, in principle valid for weak interactions, holds even in the presence of strong interactions in the ladder geometry. 

This supports the effective model, the attractive Hubbard model in the bonding band, which can also be used to predict the occurrence of superconducting phases with unconventional order in higher-dimensional geometries. In particular, in bilayer systems \cite{LanataFabrizio2009, MaierScalapino2011, GallKoehl2021, SunWang2023, GuptaHawthorn2025}
the density-assisted hopping between the planes would lead to an effective attractive two-dimensional Hubbard model for the bonding band, which is known to show superconductivity \cite{ScalettarDagotto1989, GunnarssonToschi2015, QinGull2022}. 
The $s$-wave superconductivity in the bonding band would lead to a non-local pairing between the bilayers. 
The two-band bilayer Hubbard model is relevant for the description of the nickelate La$_3$Ni$_2$O$_7$ \cite{SunWang2023, GuptaHawthorn2025, MaierDagotto2026}, where the interlayer hopping is the largest and the NiO planes form a Lieb lattice, analogous to the CuO planes in cuprates. It would be worthwhile to consider if downfolding that model to a single band model can give rise to a significant density assisted hopping. 

This work calls for more research on the Hubbard model with density-assisted hopping terms and in particular an experimental realization as for example possible in ultracold atomic gases.

% THE FOLLOWING COMMAND FORCES THE WORD COUNT TO IGNORE EVERZTING FROM THIS POINT ON!
%TC:ignore

\emph{Acknowledgments:}
We thank J.B. Bernier, T. Giamarchi and S. White for fruitful discussions.
We acknowledge support by the Deutsche Forschungsgemeinschaft (DFG, German Research Foundation) under Project No.~277625399-TRR 185 OSCAR (``Open System Control of Atomic and Photonic Matter'', B4), No.~277146847-CRC 1238 (``Control and dynamics of quantum materials'', C05), No.~511713970-CRC 1639 NuMeriQS (``Numerical Methods for Dynamics and Structure Formation in Quantum Systems''), and under Germany’s Excellence Strategy – Cluster of Excellence Matter and Light for Quantum Computing (ML4Q) EXC 2004/1 – 390534769. 
This research was supported in part by the National Science Foundation under Grants No.~NSF PHY-1748958 and PHY-2309135. R.C. acknowledges support by the PNRR MUR project PE0000023 “National Quantum Science and Technology Institute” (NQSTI), through the cascade funding projects TOPQIN and SPUNTO.

\textit{Data and materials availability:} Processed data used in the generation of main and supplementary figures are available in Zenodo with the identifier \cite{LisandriniKollath2026_dataset}. 
% 10.5281/zenodo.17295289

%

%%%%%%%%%%%%%%%%%%%%%%%%%%%%%%%%%%%%

\appendix

\section{Extraction of the central charge}

\label{SEC_CCFIT}

\begin{figure}
 \centering
 \includegraphics[width=0.48\columnwidth]{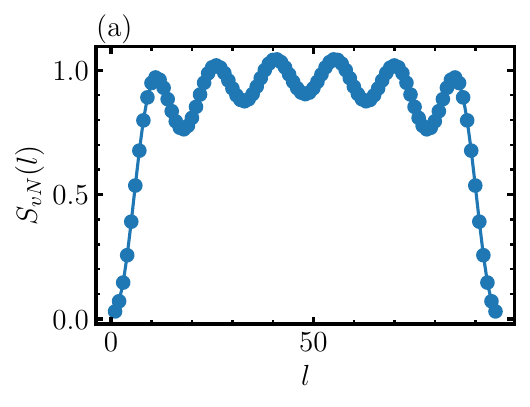}
 \includegraphics[width=0.48\columnwidth]{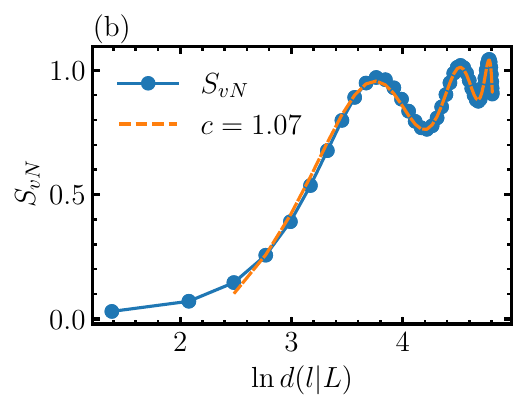}
 \caption{ Entanglement entropy for a ladder with $L=96$ rungs, as (a) as a function of the bond at which the system is bisected, and (b) as a function of the logarithmic conformal distance, $\ln d(l|L) $. Orange lines on panel (b) correspond to our fitting function in Eq.~\ref{EQ_ccfit}. The data corresponds to the case with $U=4t_\parallel$, $\tilde{t}_\perp=3t_\parallel$, $n=0.9375$ and $c_n=0.6$.
 \label{FIG_ccfit}}
 \end{figure}

An approach to identify correlated phases in one dimension is the study of the von Neumann entropy between two complementary sections of the system.  In this supplement, we describe how we extract the value of the central charge from the von Neumann entropy. 
The value of the entropy in an open system depends on the size of the system $L$, the bisected bond $l$, but most fundamentally on the number of gapless modes present in the system \cite{CalabreseCardy2004}. The entropy of a conformally invariant one-dimensional system with open boundary conditions is
\begin{equation*}
S_\text{vN}(l) = \frac{c}{6} \ln d(l|L) +\log g+ c_1,
\end{equation*}
where $d(l|L) = \frac{2L}{\pi} \sin\left(\frac{\pi l}{L} \right)$ is the conformal distance, $c$ is the central charge of the system, $\log g$ is the boundary entropy \cite{AffleckLudwig1991} and $c_1$ is a non-universal constant. 
An additional oscillating term beyond the prediction of conformal field theory is present in many systems \cite{LaflorencieAffleck2006, LegezaNoack2007, RouxAzaria2009}.
In panel (a) of Fig.~\ref{FIG_ccfit} we show the results for the entropy as a function of the bisected bond $l$ for the Hamiltonian $H$ in a ladder with $L=96$ rungs and open boundary conditions. 
The bisections are taken cutting only along the horizontal bonds ($t_\parallel$) and not cutting any rung. 
The presence of an oscillatory term is evident. 
To extract the central charge, we follow a similar approach as in Ref.~\cite{RouxAzaria2009}. We fit the function
\begin{equation}
\label{EQ_ccfit}
S_\text{vN}(l) = \frac{c}{6} \ln d(l|L) + B t(l) + A,
\end{equation}
where $t(l)= \sum_{\alpha,\sigma} \mathrm{Re} \expval{\hat{c}_{\alpha,\sigma,l}^{\dagger} \hat{c}_{\alpha,\sigma,l+1}^{\pdagger}}$ is the local kinetic energy, and $c$, $A$, $B$ are fitting parameters. 
We obtained $t(l)$ from our simulations. 

An example of the entanglement entropy as a function of the logarithmic conformal distance is plotted in panel (b) of Fig.~\ref{FIG_ccfit} with blue circles. The data correspond to the case with $U=4t_\parallel$, $\tilde{t}_\perp=3t_\parallel$, $n=0.9375$ and $c_n=0.6$. 
The orange lines in panel (b) correspond to the fitted function plotted in its fitting range. We have found that typically for the obtained results this fitting procedure including the oscillations of $t(l)$ is more robust. 

We see a good agreement of the fitting function with the data. We extract a central charge of approximately $c\simeq1.07$ for the parameters shown. 
For small $d(l|L$) (close to the boundaries), the fitting function starts deviating from the data. These are boundary effects and are omitted in the fitting.  
The value obtained in this example, namely $c\simeq1.07$, is slightly higher than the expected value, $c=1$.
This can be explained by finite-size effects. There is a finite contribution to the central charge coming from the spin gapped sector\cite{CalabreseCardy2009}. 
If both, a gapped spin mode and a gapless charge mode, are present, we will find that the entanglement entropy is $S_\text{vN} = ( \ln L +\ln \xi_{sz} )/6$, 
for open boundary conditions and $L\gg \xi_{sz}$. Therefore, the effective central charge goes as $c = 1 + \ln \xi_{sz} / \ln L$. 
Consequently, the fitted central charge is expected to decrease monotonically with increasing system size. 

\section{Correlation functions}

\begin{figure}
\centering
\includegraphics[width=0.8\columnwidth]{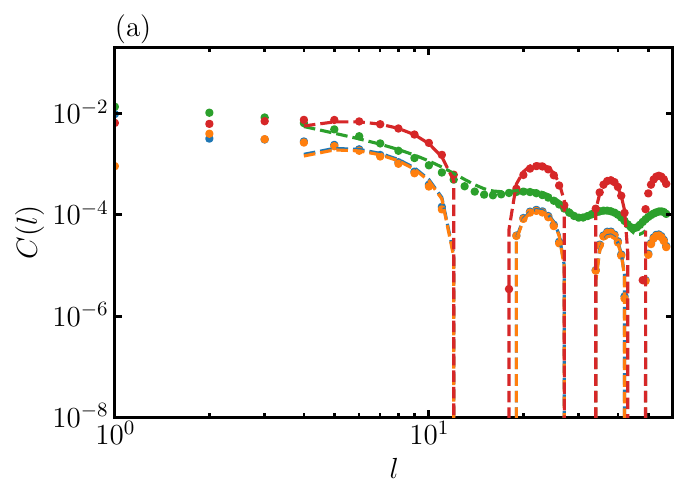} 
\includegraphics[width=0.8\columnwidth]{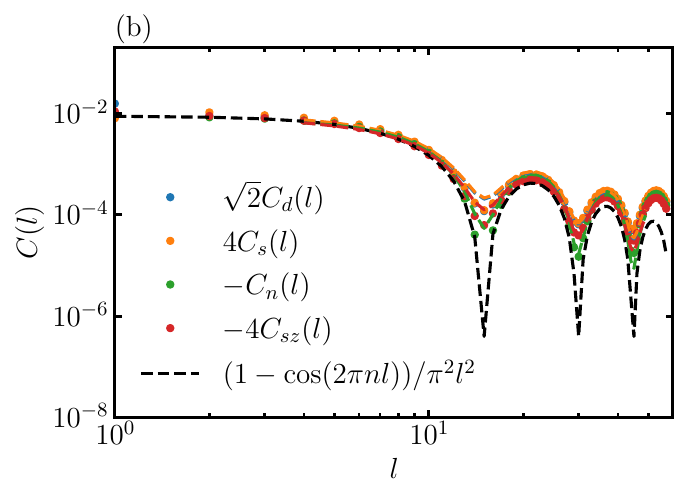}
\includegraphics[width=0.8\columnwidth]{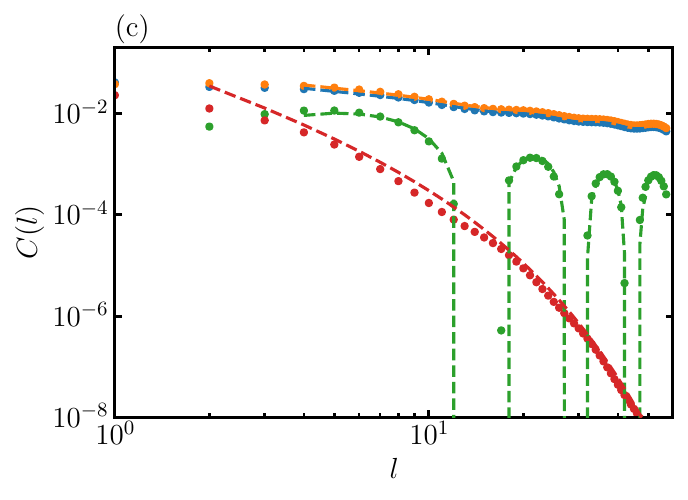} 
\caption{ Correlation functions with distance for a $L=96$ rungs ladder at $U=4t_{\parallel}$, $n=0.9375$, and $\tilde{t}_\perp=3t_{\parallel}$. Spin, density and pair correlation functions are shown. The data corresponds to a density-assisted hopping ratio with a value (a) $c_n=0.10$, (b) $c_n = 0.24$, and (c) $c_n = 0.40$. Dashed lines correspond to the fitted functions. (d) Momentum distribution for the up-spin [$n(k)$] in the bonding and antibonding band for different values of the density-assisted hopping ratio ($c_n$). 
\label{FIG_corrs} }
\end{figure}

We have numerically obtained the ground state of the system and computed the spin, $s$- and $d$-wave pair, and density correlation functions using MPS methods. In this supplement, we show examples of the decay of the correlations.

The model is symmetric under leg inversion, hence, we are going to express the correlations either in the symmetric or antisymmetric sector. 
The definition of these correlation functions are 
\begin{align*}
& C_{n}(i,j) = \expval{ \hat{n}_{j} \hat{n}_{i} } - 
  \expval{ \hat{n}_{j} } \expval{ \hat{n}_{i} }, \nonumber \\ 
& C_{sz}(i,j) = \expval{ \hat{S}^z_{j} \hat{S}^z_{i} }, \nonumber \\
& C_{s}(i,j) = \expval{ \hat{\Delta}^{\dagger}_{s, i} \hat{\Delta}^{}_{s, j} }, \nonumber \\
& C_{d}(i,j) = \expval{ \hat{\Delta}^{\dagger}_{d, i} \hat{\Delta}^{}_{d, j} }, 
\end{align*}
where the total rung density and spin operators are 
$\hat{n}_{j} = \sum_\alpha \left( \hat{n}_{\alpha, \uparrow, j} + \hat{n}_{\alpha, \downarrow, j} \right)$, and
$\hat{S}^z_{j} = \frac{1}{2} \sum_\alpha \left( \hat{n}_{\alpha, \uparrow, j} - \hat{n}_{\alpha \downarrow j} \right)$; 
and the pair creation operators are $\Delta_{s, j} = \hat{c}_{1, \uparrow, j} \hat{c}_{1, \downarrow, j} + \hat{c}_{2, \uparrow, j} \hat{c}_{2, \downarrow, j}$ 
and $\hat{\Delta}_{d, j} = \hat{c}_{1, \uparrow, j} \hat{c}_{2, \downarrow, j} - \hat{c}_{1, \downarrow, j} \hat{c}_{2, \uparrow, j}$ 
(note that these were chosen to be symmetric under leg inversion). 

To reduce the amount of boundary effects introduced by the open boundary conditions used in the MPS simulations, we averaged over different distances $l=|i-j|$, ignoring the correlation that involves sites close to the boundaries. 
We describe the way in which we process the correlation functions obtained from the simulations using the symmetric ($k_{\perp}=0$) density correlations as an example.
This reads 
\begin{equation*}
C_{n} (|j-i|) = \sum_{i,j= 1+L_b}^{L-L_b} C_{n, 0}(i,j) / N_{|i-j|}. 
\end{equation*}
where $L_b$ is the number of boundary sites discarded, and $N_{|i-j|}$ is the number of times a distance $l$ appears in the summation, $ N_{l} = 2(L - 2L_b - l)$ for $l>0$. 
We use $L_b=L/5$ except where indicated.

\textit{Bosonization predictions. }
We use the bosonization treatment of the effective Hamiltonian Eq.~\ref{EQ_Heff} in order to predict the properties of the system. The bosonized version of the Hamiltonian is given by 
\begin{eqnarray*}
H_b &=& \frac{1}{2\pi} \sum_{\nu=\rho,\sigma} 
\int dx 
\left[
u_\nu K_\nu (\partial_x \theta_\nu)^2 
+ \frac{u_\nu}{K_\nu} (\partial_x \phi_\nu)^2
\right] \\ 
&& - \frac{2g_{1\perp}}{(2\pi \alpha)^2} \cos \sqrt{8} \phi_\sigma 
\end{eqnarray*}
where $u_\nu$ are the mode velocities and $K_\nu$ the Luttinger parameters, $g_{1\perp}$ the backscattering, and $\phi_\nu$ and $\theta_\nu$ are, respectively, the bosonic field and its dual field (with $\nu = \rho,\sigma$ labeling the charge and spin sectors). 
Within this treatment one can predict the decay of the correlation functions in the different regimes. 
The charge mode $\phi_\rho$ is always gapless, but the renormalization group treatment\cite{Giamarchibook} of the quantum sine-Gordon Hamiltonian describing the spin mode $\phi_\sigma$ identifies a gapped phase (C1S0) in which $e^{i \sqrt{2} n \phi_\sigma}$ has long range order and a gapless phase (C1S1) in which $g_{1\perp} \to 0$ and $K_\sigma\to 1$  so that $e^{i\sqrt{2}\phi_\sigma}$ has only quasi-long range order. 
In the C1S0 phase (Luther-Emery), bosonization\cite{Giamarchibook} predicts that the long-distance behavior of the density and pair correlation functions is governed by a power-law decay determined by the Luttinger liquid parameter $K_\rho$, i.e., 
\begin{align}
& C_{s,d}(l,0) \simeq A_{s,d} l^{-1/K_\rho} +  B_{s,d} \cos(2\pi n l) l^{-K_\rho-1/K_\rho}
, \nonumber \\
& C_{n}(l,0) \simeq -K_\rho (\pi l)^{-2} + B_n \cos(2\pi n l) l^{-K_\rho}. 
\label{EQ_corr_C1S0}
\end{align}
where $A_{s,d}$ and $B_{s,d,n}$ are non-universal constants and $n$ is the density. 
In this regime, the spin correlation function $C_{sz}$ decays exponentially. 
In the C1S1 phase (Luttinger liquid), the spin correlation function should also decay as a power law\cite{Giamarchibook}, with logarithmic corrections coming from the renormalization flow
\begin{align*}
C_{s,d}(l,0) & \simeq A_{s,d} l^{-1 -1/K_\rho} \nonumber \\
& + B_{s,d} \cos(2\pi n_0 l) l^{-K_\rho-1/K_\rho} (\ln l)^{-3/2}
, \nonumber \\
C_{n}(l,0) & \simeq -K_\rho (\pi l)^{-2} + B_{n} \cos(2\pi n_0 l) l^{-1-K_\rho} (\ln l)^{-3/2} \nonumber \\
& + C_{n} \cos(4\pi n_0 l) l^{-4 K_\rho}
, \nonumber \\
C_{sz}(l,0) & \simeq  (2\pi l)^{-2} + B_{sz} \cos(2\pi n_0 l) l^{-1-K_\rho} (\ln l)^{1/2}. 
\end{align*}

In Fig.~\ref{FIG_corrs}(a), we show our numerical results for $t_\perp = 3t$ and $U=4t$ in the non-density-assisted case ($c_n=0$). 
In these plots, the correlation functions were scaled to collapse in the effectively non-interacting case Fig.~\ref{FIG_corrs}(b). 
As expected in panel (a), spin, density, and charge correlation functions decay as a power law compatible with the predictions from bosonization. 
Both pair correlation functions behave very similarly beyond very short distances, with the $d$-wave correlations being slightly larger in amplitude. 
For these distances, the spin and pair correlations are dominated by the oscillating term, while the density correlations are not. 
It is also shown that the decay of the pair correlations is slightly faster than the other ones, while the spin correlations are slightly dominant as expected. 

In Fig.~\ref{FIG_corrs}(c), we show our MPS results for a density-assisted hopping ratio $c_n=0.4$. 
The most evident feature is the exponential decay of the spin correlations, evidencing the opening of the spin gap. 
At the same time, the correlations belonging to the symmetric charge sector still show a power-law decay. 
Both pair correlations are clearly dominant. 
This result implies that with increasing amplitude of the density-assisted hopping term the system is not only capable of re-entering the spin-gapped regime, but also of restoring and enhancing the pair correlations. 

\begin{figure}
\centering
\includegraphics[width=\columnwidth]{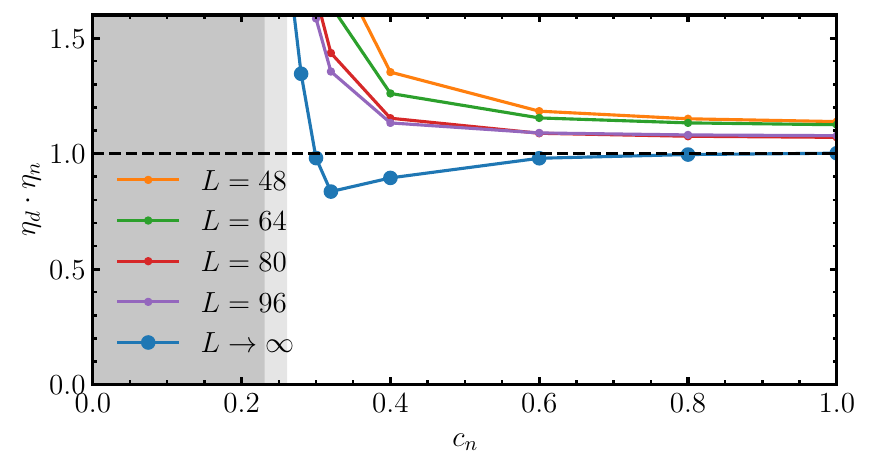} 
\caption{ Power-law exponents for the correlation functions for different system sizes as a function of $c_n$. The data corresponds to $U=4t$, $n=0.9375$, and effective perpendicular hopping is $\tilde{t}_\perp=3t$. We can see that with increasing system size, the product of the charge and pair exponents tends to one ($\eta_n \eta_d \to 1$) as expected in the thermodynamic limit. 
\label{FIG_exp_scal}}
\end{figure}

In the end of the section, we comment on the relations between the exponents extracted from the different correlation functions. We see from Eq.~\ref{EQ_corr_C1S0} that the dominating exponents are predicted to be related by $K_\rho = \eta_n = 1/\eta_d $, therefore $\eta_n \eta_d = 1 $. 
In Fig.~\ref{FIG_exp_scal}, we plot this product $\eta_n \eta_d$. We can see that finite systems are subject to some scaling, but in the thermodynamic limit the extrapolation agrees with one, that is, $\eta_n \eta_d \to 1 $ for large enough $c_n$.

\section{Fitting function}

In this section, we describe the fitting procedure to extract the correlation length from the exponentially decaying correlations and the exponents from the power-law decaying correlations.
The correlations in a gapped sector decay exponentially as 
\begin{equation*}
C(l) =  A e^{-l/\xi} l^{-\eta}
\end{equation*}
where $\eta$, $A$, and $\xi$, the correlation length, are our fitting parameters. We plot $\ln C(l)$ vs $\ln(l)$ and use a least squares regression to determine our parameters. A diverging correlation is an indication that we should fit a pure power law instead. 

The extraction of the power-law exponents is a more difficult task. 
We have tried different fitting functions and methodologies to extract the dominant exponent from these correlation functions. 
A generic power-law correlation function can be written as 
\begin{align*}
C(l) =  A l^{-\eta_{l}} +  B \cos(2\pi n_0 l) l^{-\eta_{o}}
\nonumber \\
+ \text{faster decaying terms}. 
\end{align*}
A possible approach could be to directly use this expression with $A$, $B$, $\eta_{l}$ and $\eta_{o}$ as fitting parameters. 
The most common approach in the literature (see, for example,e, \cite{GannotKivelson2020, ShenQin2023}) is to fit a single power-law function without an oscillating term, namely $Al^{-\eta}$ in a $\log$-$\log$ scale. The advantages of this approach are the simplicity and the few parameters, but it does not take into account the oscillations. 
We have taken an intermediate approach similar to the one in Ref.~\cite{SheikhanKollath2019} which we have found to work best for our case. 
To extract them from the numerically obtained correlation functions, we use the following fitting function, 
\begin{equation*}
C(l) = A l^{-\eta} + B l^{-\eta} \cos(2\pi n_0 l), 
\end{equation*}
where $\eta$, $A$, and $B$ are fitting parameters, and $n_0$ is the average density in the bulk of the system (obtained from $\expval{ n_{0 \sigma j} }$). 
We show the result of these fittings with lines in Fig.~\ref{FIG_corrs}(a-c). 
This simplified approach works well because we always find one of the following three situations:
(a) either the oscillating part is much larger than the non-oscillating part or vice versa ($A \gg B$ or $B \gg A$); 
or (b) the exponents are similar although the prefactors are fairly different ($\eta_{l} \sim \eta_{o}$). 
Situation (a) is occurring, for example, for the pair correlation in the C1S0 phase, where the non-oscillating part is clearly dominant. 
Situation (b) occurs, for example, when we are close to the transition, therefore, the exponents become similar ($K_{\rho} \to 1$). The situation (b) also occurs for the density correlations above the transition, for that case both prefactors are fairly different but the exponents are similar. 

When fitting decaying function, working with a $\log$-$\log$ scale is the standard approach. To use a logarithmic scale for the vertical axis one need to apply the absolute value to avoid negative values. Here, due to the sign change, we opted for a different approach. Staying in a linear scale can solve this problem at the cost of introducing a different one: because of the very small values of the correlation at large distances, these have too little weight in the fitting procedure. To cure this, we have rescaled the vertical axis $y$ by $y \to y x^\eta$. We have used $\eta = 1.5$, which have lead to a robust fitting procedure across the phase diagram.

\section{Pair correlation enhancement in the Luther-Emery liquid phase}

\begin{figure}
\centering
\includegraphics[width=0.45\textwidth]{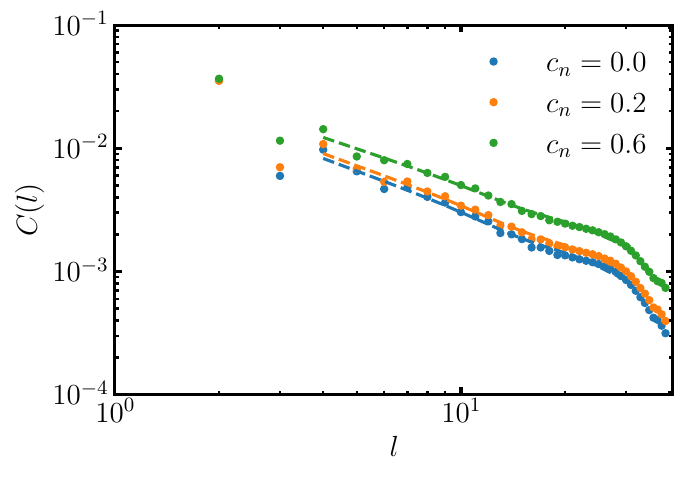} 
\caption{Pair correlation functions with distance for a $L=64$ rungs ladder at $U=4t_{\parallel}$, $n=0.9375$, and $\tilde{t}_\perp=t_{\parallel}$. Dashed lines correspond to the fitted functions. 
\label{FIG_corrsLEL}}
\end{figure}

Density-assisted hopping also enhances superconductivity in the C1S0 regime in the isotropic case ($t_\perp=t_\parallel$). This can be seen directly from the exact mapping of the density-assisted hopping term into the bonding/antibonding basis, the density-assisted hopping acts as an attraction for the lower band favoring the formation of pairs. However, since for this case, the upper and lower band are not well separated, the discussion needs to be performed with greater care.  In Fig.~\ref{FIG_corrsLEL}, we show numerical results for the full ladder model in the Luttinger liquid regime ($\tilde{t}_\perp = t_\parallel$). In this Figure, we can explicitly see that the density-assisted hopping enhances pair correlations also in this phase. Not only the magnitude of the correlation increases, but also the corresponding exponent. The fitted value at $c_n=0.0$ is $1/\eta_d = 0.90$ and increases to $1/\eta_d = 1.0$ for $c_n=0.6$. 

\section{Momentum distribution}

We define the single-particle momentum distribution as the Fourier transformation of the single-particle correlation function (taking into account open boundary conditions), this is  
\begin{equation*}
    n_{k_\perp, \sigma} \left( k \right) = \frac{1}{\sqrt{L+1}} \sum_{i,j=1}^{L} \sin{(i k)} \sin{(j k)} \langle \hat{c}^{\dagger}_{k_\perp, \sigma, i} \hat{c}^{\pdagger}_{k_\perp, \sigma, j} \rangle, 
\end{equation*}
where $k_\perp=0,\pi$ correspond to the bonding and antibonding bands, and the allowed values of the momentum are $\frac{L+1}{\pi}k = 1, 2, \dots, L+1$. 
Note that with this definition, for free fermions, the result will be a step function. We have extracted the momentum discontinuity, $Z_{k_F} = \max_j \left[ n(k_{j}) - n(k_{j+1}) \right]$ which corresponds in the thermodynamic limit to the momentum discontinuity for a non-interacting system.

\section{Scaling behavior of the momentum discontinuity \texorpdfstring{$Z_{k_F}$}{ZkF}}

In this section, we present the derivation of the dependence of $Z_{k_F}$ with system size. 
In the Luttinger liquid phase, we obtain a power-law decay with system size. 
For $x\gg \alpha$ the single particle Green's function is given by 
$$\expval{\psi^{\dagger}_{\sigma}(x) \psi^{\pdagger}_{\sigma}(0)} \sim \frac{\text{sign}(x)}{2 \pi \alpha} \left( \frac{\alpha}{|x|} \right)^{\gamma},$$ 
with $\gamma =(2+K_\rho+K_\rho^{-1})/4$ \cite{Giamarchibook}. 
Its Fourier transform gives us the momentum distribution which, in the thermodynamic limit, is 
$$ n(k) = n(k_F) - C \text{sign}(k-k_F) |k-k_F|^{\gamma-1}, $$ 
as $k \simeq k_F$. 
In finite size with periodic boundary conditions, $k$ takes values which are integer multiples of $\frac{2\pi}{L}$. In this case, the apparent discontinuity should behave as $Z_{k_F} \sim C  (L/(2\pi))^{1-\gamma}$. 

In the Luther-Emery phase, the gapped modes contribute a power law at distances shorter than the correlation length $\xi$ and an exponential decay at distance longer than the correlation length $\xi$. We approximate the single-particle Green's function as 
$$\expval{\psi^{\dagger}_{\sigma}(x) \psi^{\pdagger}_{\sigma}(0)} = \frac{\text{sign}(x)}{2 \pi \alpha} e^{-|x|/\xi} \left( \frac{\alpha}{|x|} \right)^{\gamma}.$$ 
When we compute the momentum distribution in the thermodynamic limit $n(k)$, it is of the form
$$n(k) = n(k_F) - C' \xi^{1-\gamma} f((k-k_F) \xi).$$ 
where $f$ is of class $C^{\infty}$ when $k$ is near $k_F$.
In a finite-size system with periodic boundary conditions, the apparent discontinuity is $Z_{k_F} = C' \xi^{1-\gamma} [f(-2\pi \xi/L) -f (2\pi \xi/L)]$. Therefore, for large $L \gg \xi$, the discontinuity becomes $Z_{k_F} \sim f'(0) \xi^{2-\gamma} L^{-1}$. 
Since $\gamma-1 \leq 1$, the decay of $Z_{k_F}$ is faster as a function of $L$ than in the Tomonaga-Luttinger liquid.
When $L \ll \xi$, the correlation length is capped by the size of the system and we recover the scaling of $Z_{k_F} \sim L^{1-\gamma}$. 

We have left to mention the non-interacting case; at that point, $\gamma=1$, we obtain the usual step function $n(k)=\Theta (k_F -k)$. Therefore, there is no dependence on the size of the system, which means $Z_{k_F}(L) = \text{cte}$. 

\begin{figure}
\centering
\includegraphics[width=0.49\columnwidth]{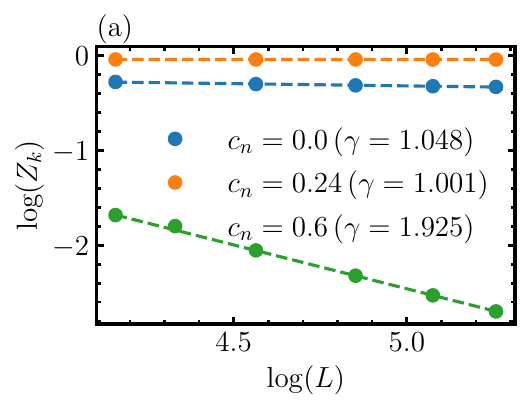} 
\includegraphics[width=0.49\columnwidth]{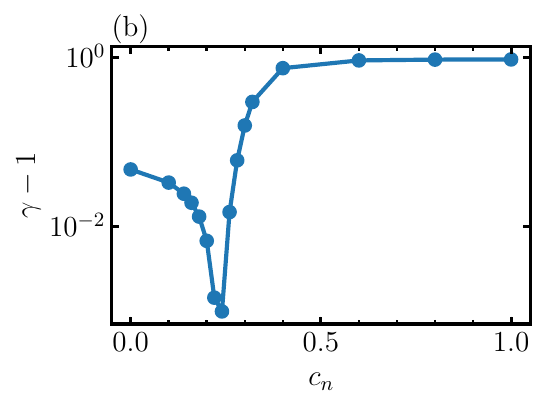} 
\caption{ Finite size behavior of the momentum discontinuity $Z_{k_F}$. The data corresponds to $U=4t_{\parallel}$, $n=0.9375$, and effective perpendicular hopping is $\tilde{t}_\perp=3t_{\parallel}$. In panel (a) we show the fittings corresponding to the quantity $\log(Z_{k_F})$ with a function $(1-\gamma) \log(L) + B$. In panel (b), we plot $\gamma$ for different values of $c_n$. 
\label{FIG_Zk_scal}}
\end{figure}

We analyze the finite-size scaling of the apparent momentum discontinuity $Z_{k_F}$ for our numerical result. In Fig.~\ref{FIG_Zk_scal} we show its behavior for different values of $c_n$. We plot in panel (a) for selected values of $c_n$, $Z_{k_F}$ as a function of the system size in log-log scale. 
We can see that, as expected, all of them follow a power law $Al^{1-\gamma}$. 
In the Luther-Emery phase, we see the strongest system size effects. 
We present these results in panel (b), where $\gamma-1$ is plotted as a function of $c_n$. Inside the Luther-Emery phase ($c_n \gtrsim 0.23$), we see that the value of the exponent approximates to two ($\gamma \to 2$). For example, for $c_n=0.6$ the decaying exponent is $\gamma \simeq 1.925(5)$. 
On the other side of the transition, in the Luttinger liquid phase ($c_n \lesssim 0.23$), the decaying is much slower than on the attractive side, but still decaying as a power law. 
For example, for $c_n=0.0$, one can see a slow decaying behavior with $\gamma \simeq 1.048(2)$. 
In the vicinity of the transition, we see that this exponent becomes almost zero, showing almost no finite-size effects. As an example, we have $\gamma \simeq 1.00099(5)$ for $c_n=0.0$. This point is in the near vicinity of the effectively non-interacting point, and it shows a sharp step function in its momentum distribution [shown in Fig.~\ref{FIG_cc_gap_Zk}(b) in the main text].

\section{Phase diagram for different parameters}

\begin{figure*}
\centering
\includegraphics[width=0.6\columnwidth]{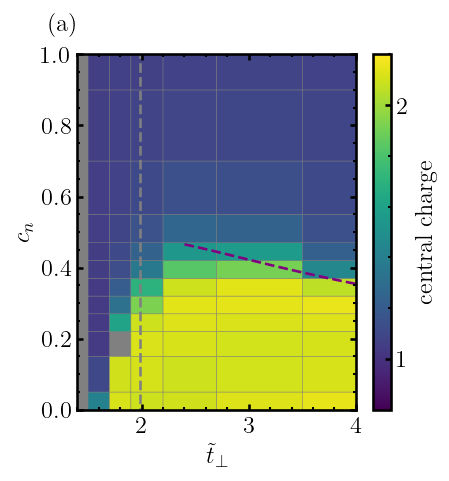}
\includegraphics[width=0.6\columnwidth]{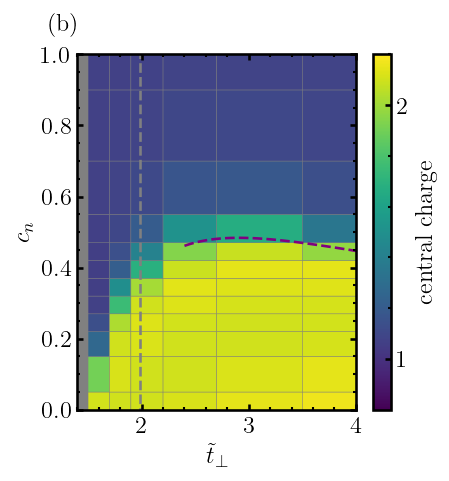}
\includegraphics[width=0.6\columnwidth]{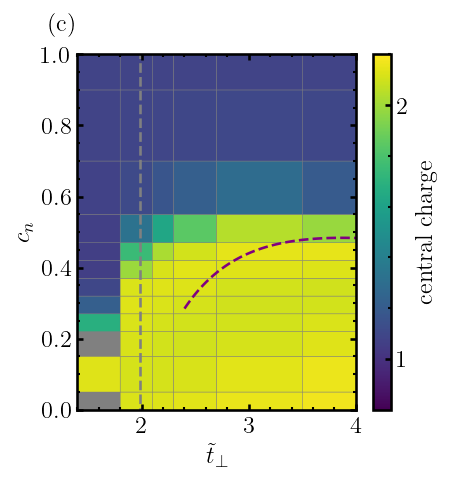}
\includegraphics[width=0.6\columnwidth]{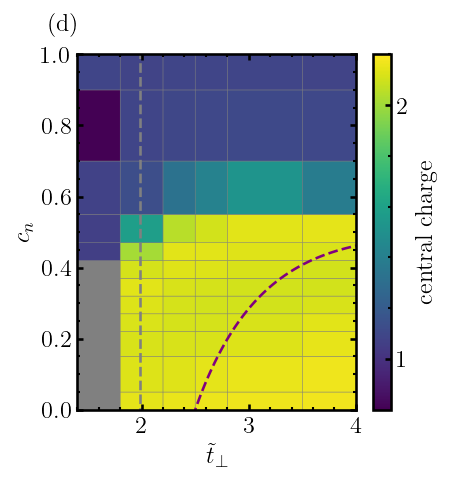}
\includegraphics[width=0.6\columnwidth]{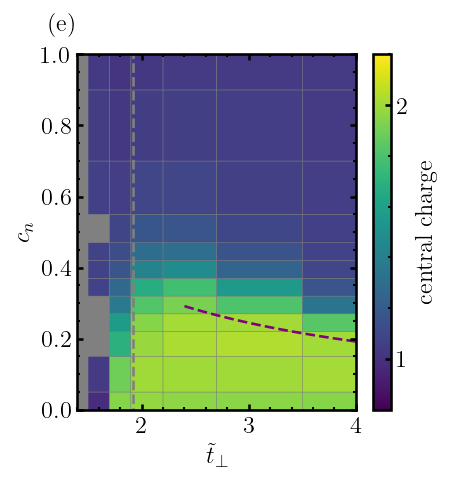}
\includegraphics[width=0.6\columnwidth]{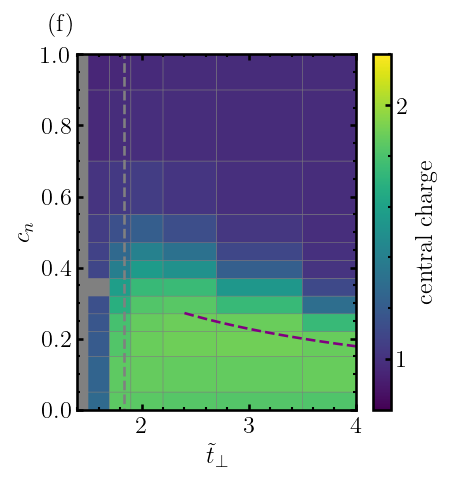}
\caption{ The value of the central charge extracted from MPS simulations on $64-$rungs systems. The purple dashed lines correspond to the values of the transition predicted by the effective model. The corresponding simulation parameters are (a) $n=0.9375$, $U=8t_{\parallel}$, (b) $n=0.9375$, $U=12t_{\parallel}$, (c) $n=0.9375$, $U=16t_{\parallel}$, (d) $n=0.9375$, $U=20t_{\parallel}$, (e) $n=0.875$, $U=4t_{\parallel}$, and (f) $n=0.8125$, $U=4t_{\parallel}$. 
\label{FIG_centralcharge_more}}
\end{figure*}

\begin{figure*}
\centering
\includegraphics[width=0.65\columnwidth]{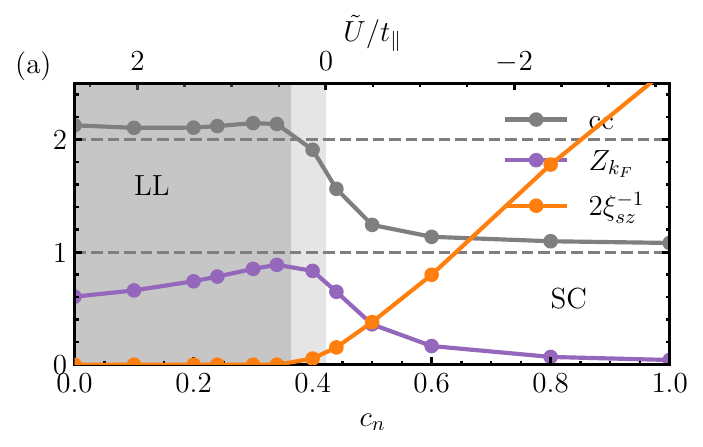}
\includegraphics[width=0.65\columnwidth]{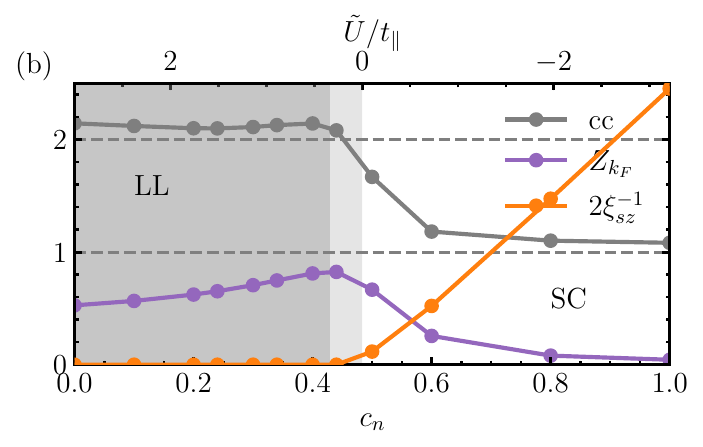}
\includegraphics[width=0.65\columnwidth]{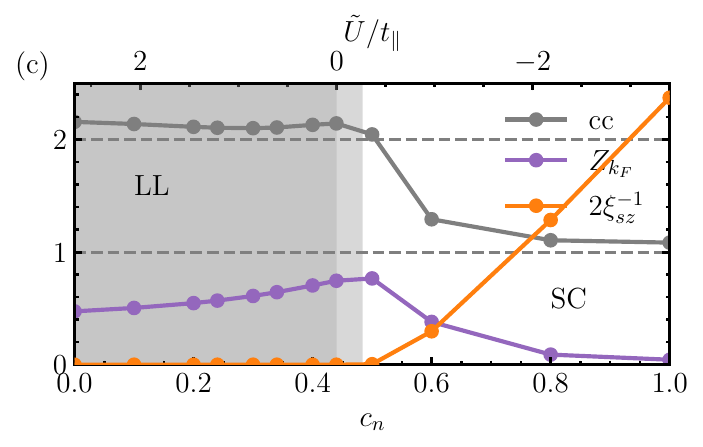}
\includegraphics[width=0.65\columnwidth]{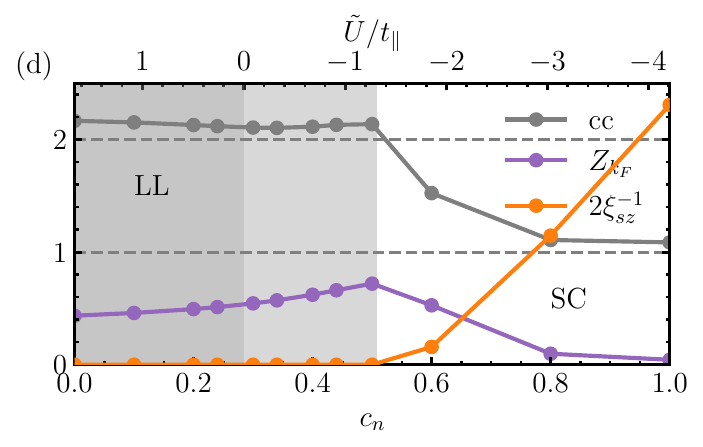}
\includegraphics[width=0.65\columnwidth]{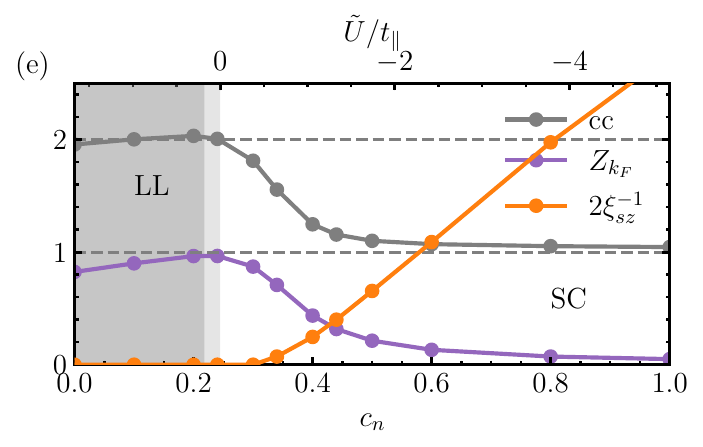}
\includegraphics[width=0.65\columnwidth]{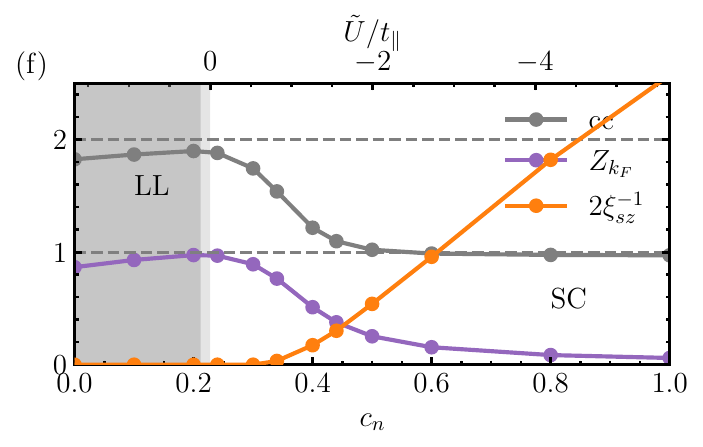}
\caption{ Transition at $\tilde{t}_\perp=3t_\parallel$ for a $64$-rung ladder. (a) The gray and purple plots correspond to the central charge and the discontinuity of the momentum distribution $Z_{k_F}$, respectively. The orange plot corresponds to the inverse correlation length. The corresponding simulation parameters are (a) $n=0.9375$, $U=8t_{\parallel}$, (b) $n=0.9375$, $U=12t_{\parallel}$, (c) $n=0.9375$, $U=16t_{\parallel}$, (d) $n=0.9375$, $U=20t_{\parallel}$, (e) $n=0.875$, $U=4t_{\parallel}$, and (f) $n=0.8125$, $U=4t_{\parallel}$. 
\label{FIG_ccgapZk_more}}
\end{figure*}

For the main part of this work, we have chosen to show the data corresponding to a value of the Coulomb interaction $U=4t_\parallel$ and density $n = 15/16 = 0.9375$, but the validity of the results is much wider than that. 
In Figs.~\ref{FIG_centralcharge_more} and \ref{FIG_ccgapZk_more}, we present results for different interaction strengths ($U=8t_\parallel$, $12t_\parallel$, $16t_\parallel$, and $20t_\parallel$), and densities ($n=14/16=0.875$ and $13/16=0.8125$). 
The purple dashed lines in Fig.~\ref{FIG_centralcharge_more} correspond to the transition predicted by our effective model. Although the effective model is, in principle, valid for small values of $U$, the agreement seems to be very good up to large values of the interaction such as $U=12t_\parallel$ [Fig.~\ref{FIG_centralcharge_more}(a-b)]. 
For larger values of the interaction, $U=16t_\parallel$ and $20t_\parallel$, the prediction of the critical $c_n$ is not very precise. 
Fermions occupy the antibonding band, and the interband terms in Eq.~\ref{EQ_U_AB} must be taken into account. 
Despite this fact, in Fig.~\ref{FIG_ccgapZk_more}, we see that the key features discussed in the main text at the transition (the opening of the spin gap, the reduction of the central charge from $2$ to $1$, and a maximum on the discontinuity $Z_k$) remain the same even for very large values of the Coulomb interaction ($U=20t_\parallel$). 
Note that the maximum value of the discontinuity of the momentum distribution ($Z_k$) is reduced as a consequence of the reduced occupation of the lower band. 

We have also probed the stability of the transition for different values of the density the results at $U=4t_\parallel$ for densities $n=14/16=0.875$ and $13/16=0.8125$ are presented in \ref{FIG_centralcharge_more} and \ref{FIG_ccgapZk_more} in panels (e) and (f), respectively. 
In these cases, the transition predicted by the effective model agrees with the results found numerically, showing that the transition generically is stable in the studied parameter space. 


%TC:endignore
\end{document}